\documentclass[aps,twocolumn,prc,superscriptaddress,noshowpacs,nofootinbib,noshowkeys,floatfix]{revtex4-2}
\usepackage[dvips]{graphics,graphicx}
\usepackage[colorlinks=true,linktocpage=true,linkcolor=blue,citecolor=blue]{hyperref}
\usepackage[usenames,dvipsnames]{color}
\usepackage{amsmath, amssymb}
\usepackage{multirow}
\usepackage{longtable}
\usepackage{color}
\usepackage{xcolor}
\usepackage{xspace}
\usepackage{cleveref}
\usepackage[normalem]{ulem}  

\renewcommand\sout{\bgroup \color{blue} \ULdepth=-.5ex \ULset}

\newcommand{\sqsn}{$\sqrt{s_{\rm{NN}}}$}
\newcommand{\sqs}{$\sqrt{s}$}
\newcommand{\pt}{$p_{\rm{T}}$\xspace}
\newcommand{\hf}{heavy flavor\xspace}
\newcommand{\hfs}{heavy flavors\xspace}
\newcommand{\pta}{$p_{\rm{T}}^{\rm assoc}$\xspace}

\newcommand{\pte}{$p_{\rm{T}}^{\rm e}$\xspace}
\newcommand{\ptd}{$p_{\rm{T}}^{\rm D}$\xspace}

\newcommand{\gev}{GeV$/c$\xspace}
\newcommand{\delphi}{$~\Delta \varphi$~}
\newcommand{\deleta}{$~\Delta \eta~$}
\newcommand{\bef}{\begin{figure}}
\newcommand{\eef}{\end{figure}}
\newcommand{\bc}{\begin{center}}
\newcommand{\ec}{\end{center}}

\newcommand{\be}{\begin{equation}}
\newcommand{\ee}{\end{equation}}
\newcommand{\bea}{\begin{eqnarray}}
\newcommand{\eea}{\end{eqnarray}}

\begin{document}

\title{{\Large Jet fragmentation via azimuthal angular correlations of heavy flavor decay electrons in pp, p--Pb, and Pb--Pb collisions using PYTHIA8+Angantyr}}

\author{Ravindra Singh,\footnote{Corresponding author.}}
\email{ravirathore.physics@gmail.com}
\author{Yoshini Bailung}
\author{Sumit Kumar Kundu}
\author{Ankhi Roy}
\email{ankhi@iiti.ac.in}
\affiliation{Department of Physics, School of Basic Sciences, Indian Institute of Technology Indore, Simrol, Indore 453552, India}

\begin{abstract}
Measurements in heavy flavor azimuthal angular correlation provide insight into the production, propagation, and hadronization of heavy flavor jets in ultra-relativistic hadronic and heavy-ion collisions. These measurements across different colliding systems, like p--A and A--A, help us isolate the possible modification in particle production due to cold nuclear matter (CNM) effects and the formation of Quark-Gluon Plasma (QGP), respectively. Jet correlation studies give direct access to the initial parton dynamics produced in these collisions. 

This article studies the azimuthal angular correlations of electrons from heavy flavor hadron decays in pp, p--Pb, and Pb--Pb collisions at \sqsn = 5.02 TeV using PYTHIA8+Angantyr. We study the production of heavy flavor jets with different parton level processes, including multi-parton interactions, different color reconnection prescriptions, and initial and final state radiation processes. In addition, we add the hadron-level processes, i.e., Bose-Einstein and rescattering effects, to quantify the effect due to these processes. The heavy flavor electron correlations are calculated in the different trigger and associated \pt intervals to characterize the impact of hard and soft scattering in the various colliding systems. The yields and the sigmas associated with the near-side (NS) and away-side (AS) correlation peaks are calculated and studied as a function of associated \pt for different trigger \pt ranges.


\end{abstract}

\maketitle
\section{I. Introduction}
\label{intro}
Heavy nuclei are collided at the high center of mass energies to produce a primordial state of matter, the Quark-Gluon Plasma (QGP), which is a deconfined state of quarks and gluons. The likes of experiments such as the Relativistic Heavy-Ion Collider (RHIC) at BNL, USA, and the Large Hadron Collider (LHC) at CERN, Geneva, Switzerland, serve the purpose of studying this state of matter, and unraveling its properties~\cite{Busza:2018rrf, STAR:2005gfr, ALICE:2008ngc}. These collider experiments aim to probe the strongly interacting matter phase diagram, which is based on quantum chromodynamics (QCD)~\cite{Gao:2020qsj,Tawfik:2016cot,Satz:2013xja}. In very high energy regimes, measurements in hadronic (or pp) collisions provide the stringent test for perturbative QCD (pQCD) calculations and serve as a baseline for heavy-ion measurements~\cite{Collins:1989gx,Catani:1996vz}. In p--Pb systems, cold nuclear matter (CNM) effects come into play, which may modify the hadronization mechanisms and the yields of identified particles~~\cite{Arleo:2012rs, LHCb:2013gmv}. Heavy-ion collisions (Pb--Pb or Au--Au) measure the properties and dynamics of the possible QGP formed~~\cite{STAR:2005gfr, ALICE:2008ngc}.

In ultra-relativistic heavy-ion and hadronic collisions, the initial hard scatterings result in a collimated beam of high momentum (\pt) partons. These partons fragment to produce a cluster of particles, known as a jet~~\cite{CMS:2016lmd,ATLAS:2012tjt,ALICE:2017nij}. The study of high-\pt jets establishes how partons fragment into various particles and reveal the effects of their interaction with the medium. Initial hard scatterings in pp, p--Pb, and Pb--Pb collisions also lead to the production of \hfs, namely charm (c) and beauty (b)~~\cite{ALICE:2016yta,ALICE:2012acz}. Their early production is accredited to their large mass and allows them to traverse and interact with the partons in the produced hot QCD matter produced. The production cross-section of these heavy quarks can be calculated using the factorization theorem
\begin{align}
d\sigma_{AB\rightarrow C}^{\rm{hard}} = \Sigma_{a,b} f_{a/A} (x_a,Q^2) \otimes f_{b/B} (x_b,Q^2) \otimes\\
\nonumber d\sigma_{a,b\rightarrow c}^{\rm{hard}} (x_a,x_b,Q^2) \otimes D_{c\rightarrow C}(z,Q^2)
\end{align}

where, $f_{a/A} (x_a,Q^2)$ and $f_{b/B} (x_b,Q^2)$ are the parton distribution functions which give the probability of finding parton "a"(b) inside the particle "A"(B) for given x (fraction of particle momentum taken by parton) and factorization scale ($Q^2$), $d\sigma_{ab\rightarrow cX}^{\rm{hard}} (x_a,x_b,Q^2)$ is the partonic hard scattering cross-section, and $D_{c\rightarrow C}(z,Q^2)$ is the fragmentation function of the produced parton (particle).

The \hf hadron production is sensitive to the charm and bottom fragmentation functions and to the hadronization mechanisms of these \hf hadrons~\cite{Norrbin:2000zc,Faggin:2021uwx}. These heavy quarks hadronize on a shorter time scale as they traverse the medium. This phenomenon can lead to a modification in the fragmentation function of the heavy quarks. In order to quantify the medium effects, studies of high-\pt jet fragmentation are done via angular correlations of heavy flavor particles in heavy-ion collisions~~\cite{ALICE:2021kpy,Singh:2021ppc}. Azimuthal angular correlation study is an effective tool for studying jet events. A jet event can consist of a single jet, the particles from which will produce a large correlation at \delphi = 0, or a back-to-back di-jet in which the particles will produce a correlation at \delphi = $\pi$. The correlation function is obtained by correlating each trigger particle with the associated charged particle. These correlations appear as peaks in a \delphi distribution, generally known as the $``$near-side'' (\delphi = 0) and $``$away-side'' (\delphi = $\pi$) peaks.

In a pp collision, more than one distinct hard-parton interaction can occur, and proton remnants can also scatter again on each other. Such processes are called multi-parton interactions (MPI) and are responsible for the production of a large fraction of the particles. The MPI implementation used in PYTHIA8~~\cite{Sjostrand:2007gs} (which also drives the MPI process in POWHEG+PYTHIA8 simulations~~\cite{Frixione:2007vw}), charm-quark production can occur not only from the first (hardest) hard scattering but also from hard processes in the various MPI occurring in the collisions, ordered with decreasing hardness. There is also some correlation between FSR+ISR and MPI processes since initial- and final-state radiations are generated from all the parton interactions occurring in the collision and are thus enhanced in the presence of MPI.
The recent measurement of angular correlations between D mesons and charged particles by the STAR collaboration shows a significant modification of the near-side peak width and associated yield, which increases from peripheral to central Au–Au collisions~~\cite{STAR:2019qbf}. Similar measurements were later carried out by LHC, which investigated the possible modifications in jet properties due to the medium effects~~\cite{Cao:2020wlm}. The measurements show suppression for the away-side peak, suggesting energy loss of the recoil-jet parton traversing through the medium. The amount of suppression can be quantified by the near- and away-side yield ratios taken for p--Pb and Pb--Pb systems over pp where medium effects are not present. We inspect the contribution of MPI and various CR phenomena with PYTHIA8+Angantyr~~\cite{Bierlich:2018xfw} in the regime of perturbative QCD.     

In this article, the heavy-flavor hadron decay electrons (c,b$\rightarrow e$) are used to study the parton shower of heavy quarks. It will contribute to a better understanding of heavy flavour parton showers and offer predictions for measurements of the heavy-flavor correlation. This study is important from the perspective of experimental measurements at high energies in heavy flavor correlations, which are currently available only for charm mesons. By varying the trigger and associated particle \pt, this work aims to investigate how soft and hard fragmentation showers interplay. The correlation peaks in this article are described using a novel fitting function (von Mises). As the BLC tunes increase the peak amplitude for baryon-tagged correlation, predictions from the new color reconnection (BLC) tunes are compared to the default (Monash) ones to see the behavior of fragmentation functions in the presence of baryon decay electrons. Further, the effect of partonic and hadronic level processes on heavy flavor jet fragmentation is studied.

The paper is organized as follows; in section~\ref{evt_gen}, the details of event generation and analysis methodology are discussed. Section~\ref{result} presents the results and discussions. Finally, in section~\ref{sum}, we summarize our findings.

\section{Event generation and Analysis methodology}
\label{evt_gen}
PYTHIA event generator is used comprehensibly to study proton-proton, proton-lepton, proton-nucleus, and nucleus-nucleus collisions. It provides an abundance of processes and tunes to choose and implement as per the physics involved in the study. The collisions between nuclei are done with the inclusion of Angantyr with PYTHIA8~~\cite{Sjostrand:2007gs,Bierlich:2018xfw}. PYTHIA8+Angantyr helps produce a dense medium-free state of partons that further hadronize to form the final state particles.
PYTHIA includes multi-parton interactions (MPI), Initial-State Radiations (ISR) and
Final-state Radiations (FSR) implementing hard and soft parton scattering
processes. The high \pt partons form showers or jets, which undergo fragmentation processes and hadronize according to the Lund string fragmentation model. Hadronization follows the Color Reconnection (CR) mechanism between the partons, which is carried out by rearranging the strings between them. By doing so, the total string length can be modified, which is sensitive to the hadronization process. The partons hadronize when the string length is small enough after subsequent creation of light quark-antiquark pairs~~\cite{Maldonado-Cervantes:2014tva}. MPI and CR phenomena in PYTHIA contribute significantly to the particle production mechanism, as seen in the charged-particle multiplicity distributions~~\cite{Singh:2021edu,ALICE:2017pcy}

\begin{figure}
    \includegraphics[scale = 0.42]{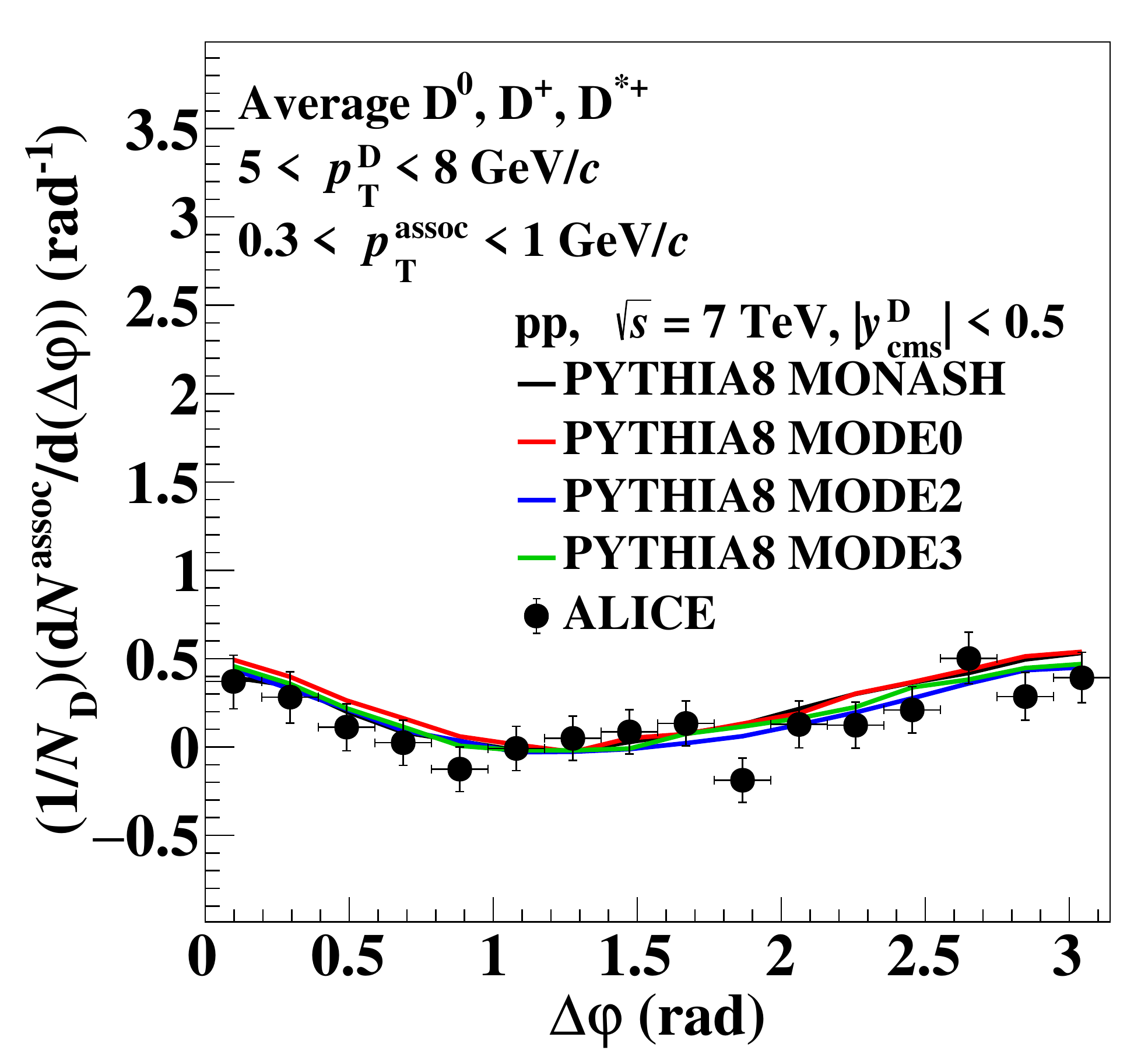}
    \caption{Comparison of average D-meson azimuthal-correlation distribution at mid-rapidity with PYTHIA8 Monash for trigger \ptd range 5 $<$ \ptd $<$ 8 \gev and \pta range 0.3 $<$ \ptd $<$ 1 \gev in pp collisions at \sqs $=$ 7 TeV.}
    \label{fig:delphi1}
\end{figure}

\begin{figure}
    \includegraphics[scale = 0.425]{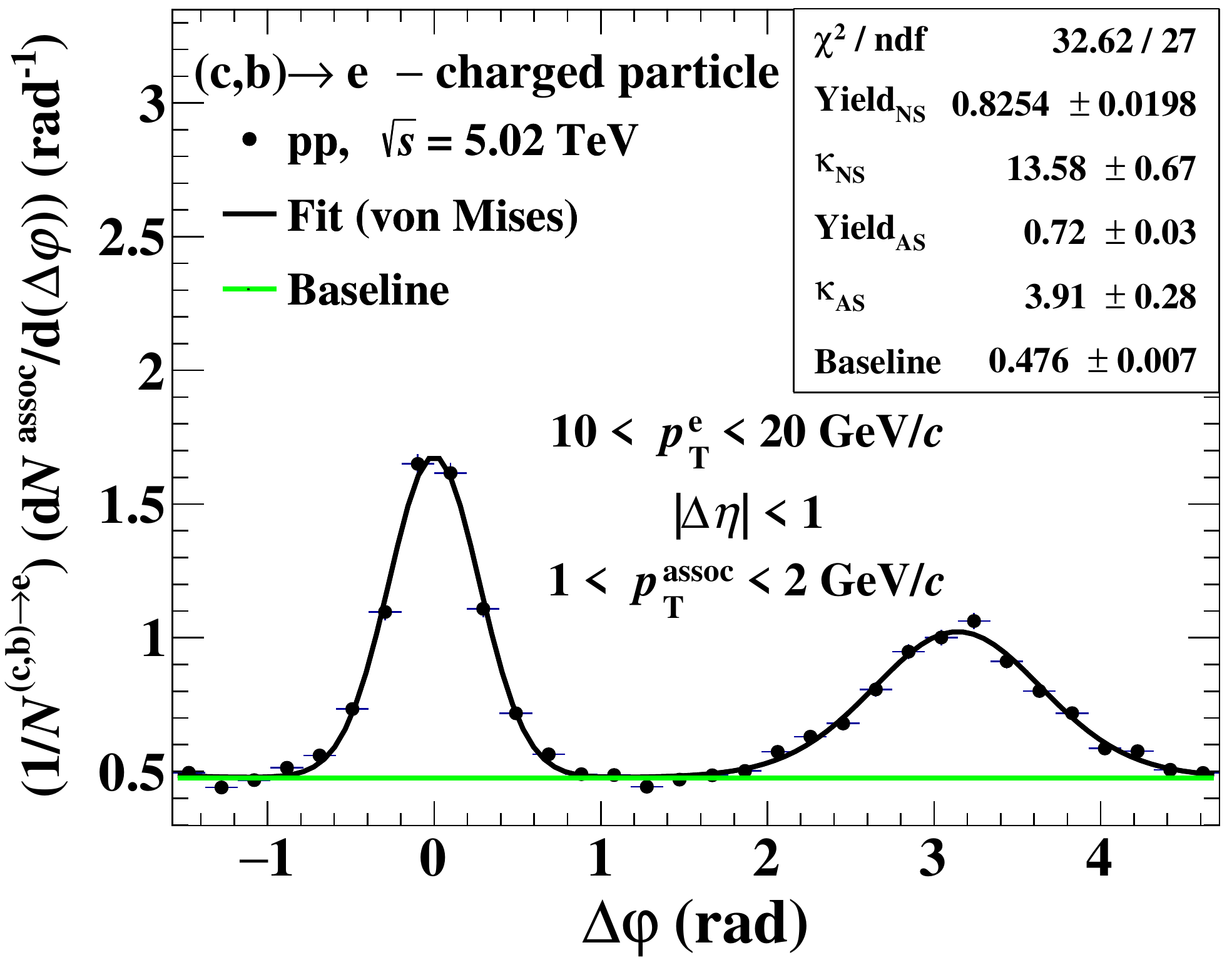}
    \caption{(Color online) The azimuthal-correlation distribution (\delphi) fitted with the von Mises function is shown for trigger \pte range $ 10 < p_{T}^{e} < 20$ GeV$/c$ and for associated $p_{\rm T}$ range $1 < p_{T}^{e} < 2$ \gev in pp collisions at \sqs = 5.02 TeV.}
    \label{fig:fitdelphi}
\end{figure}

\begin{figure*}
\includegraphics[scale=0.40]{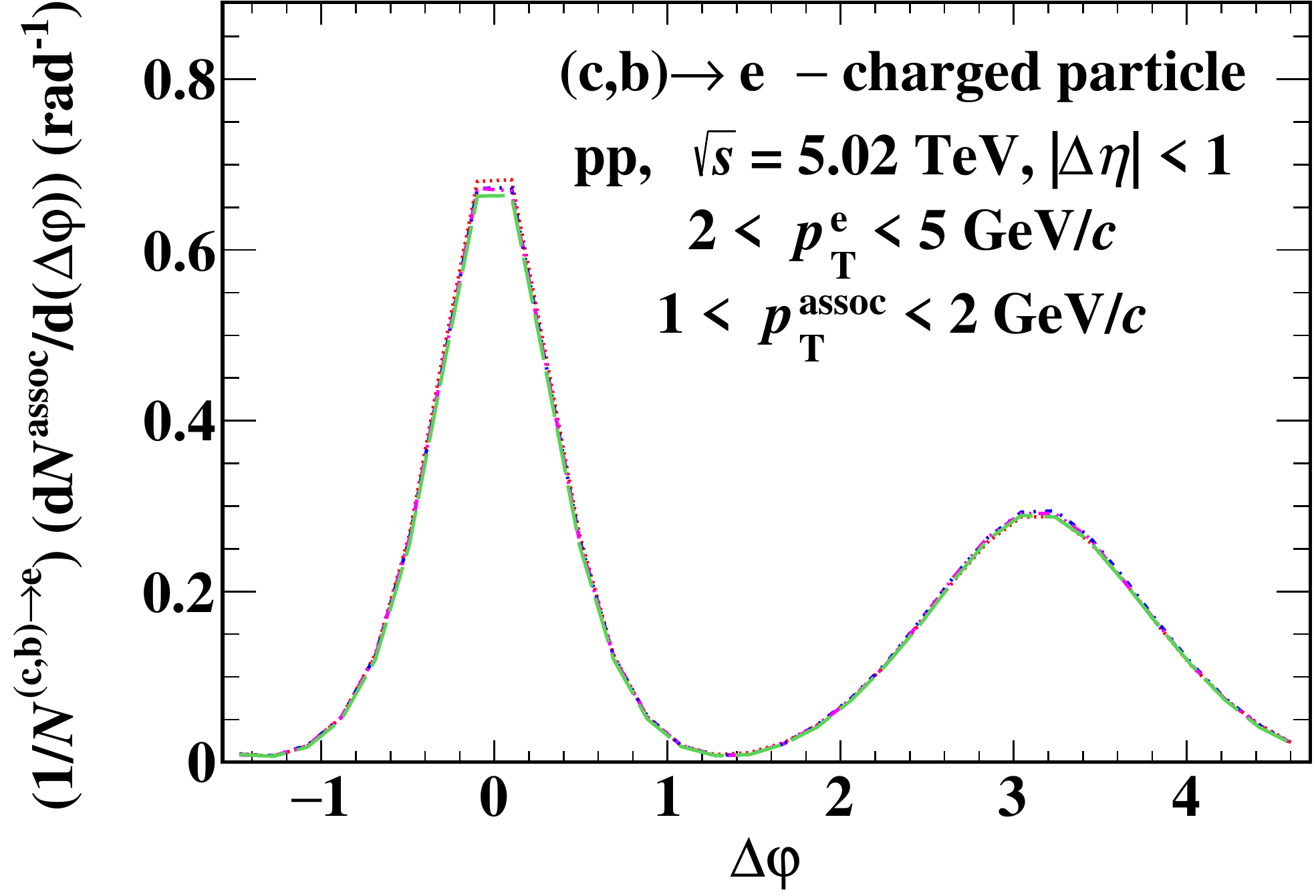}
\includegraphics[scale=0.40]{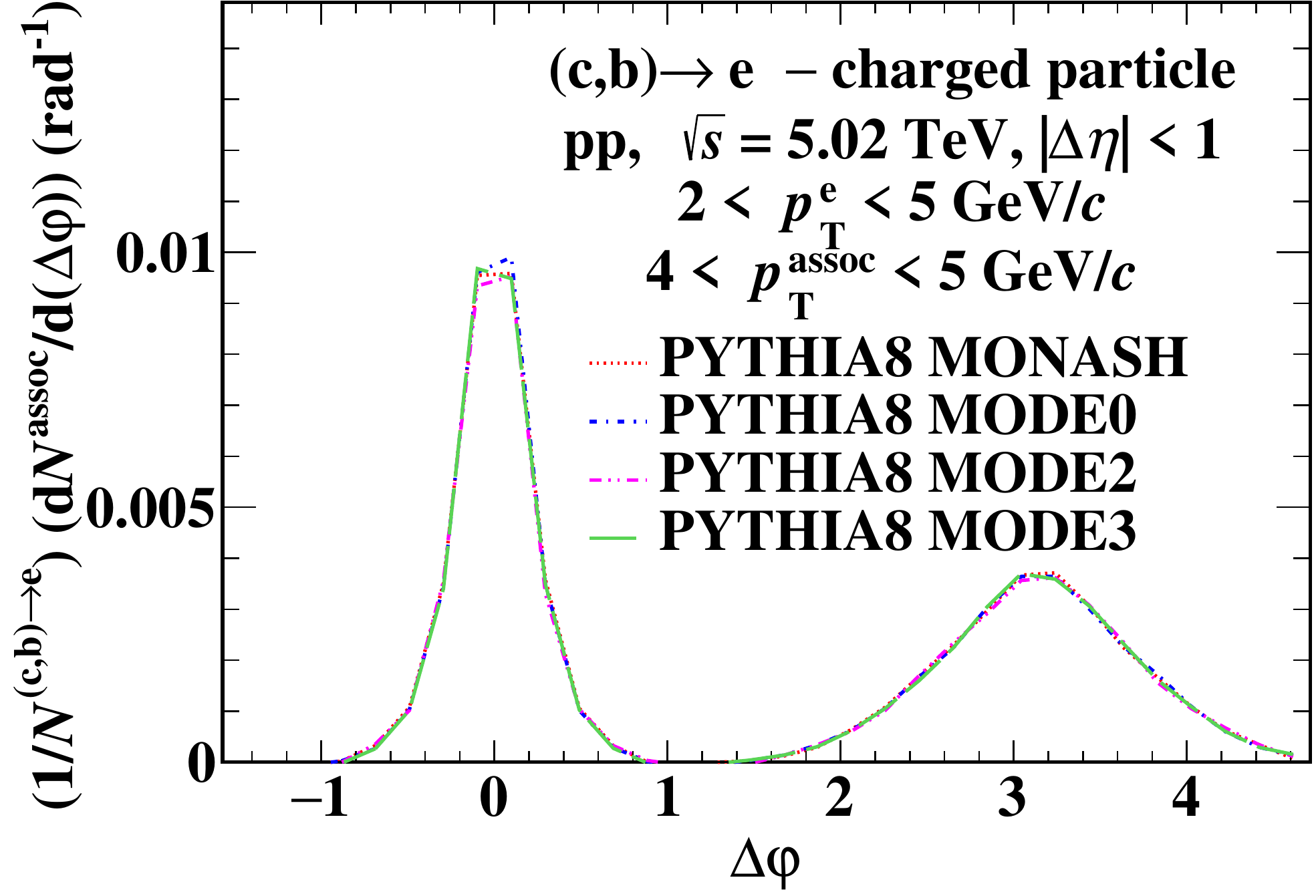}\\
\includegraphics[scale=0.40]{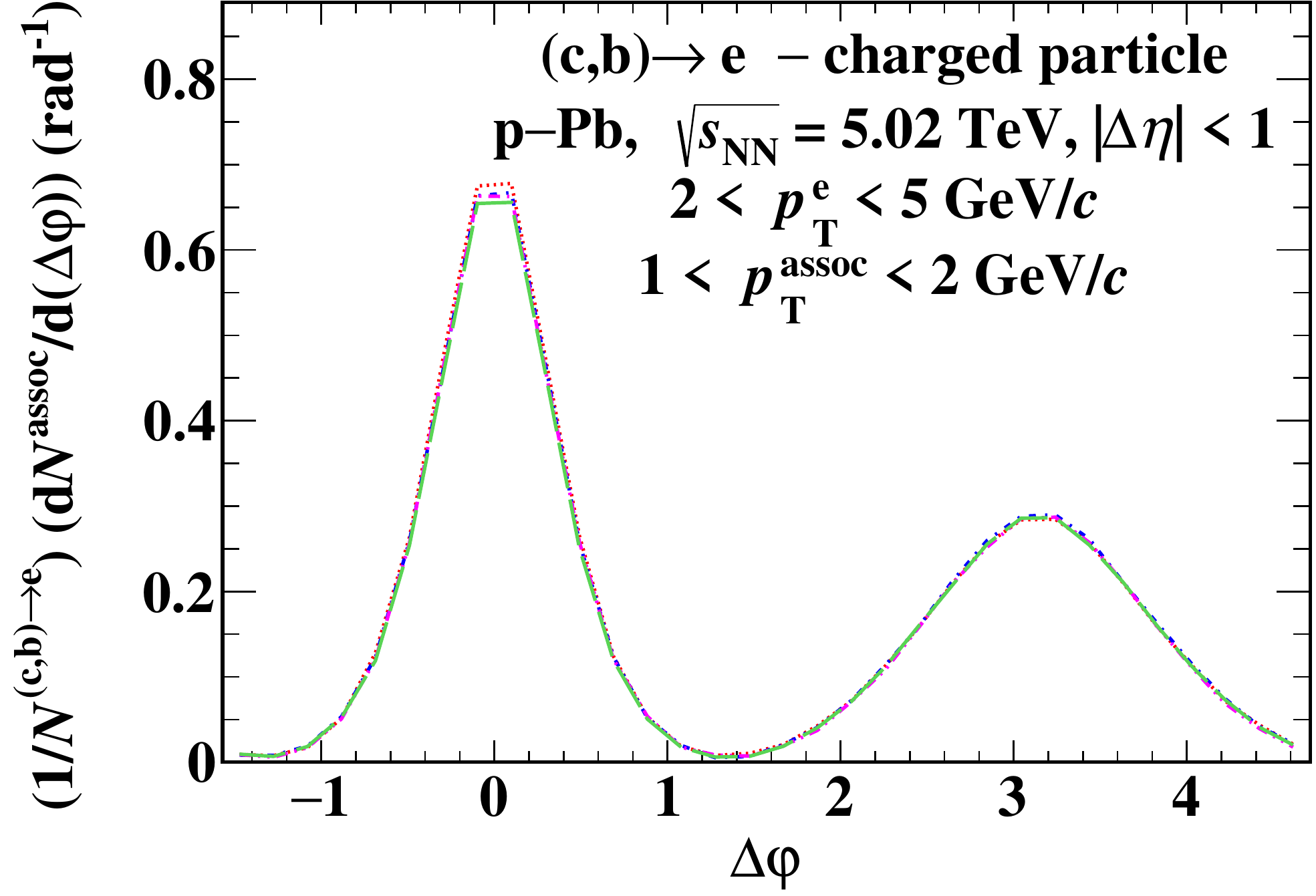}
\includegraphics[scale=0.40]{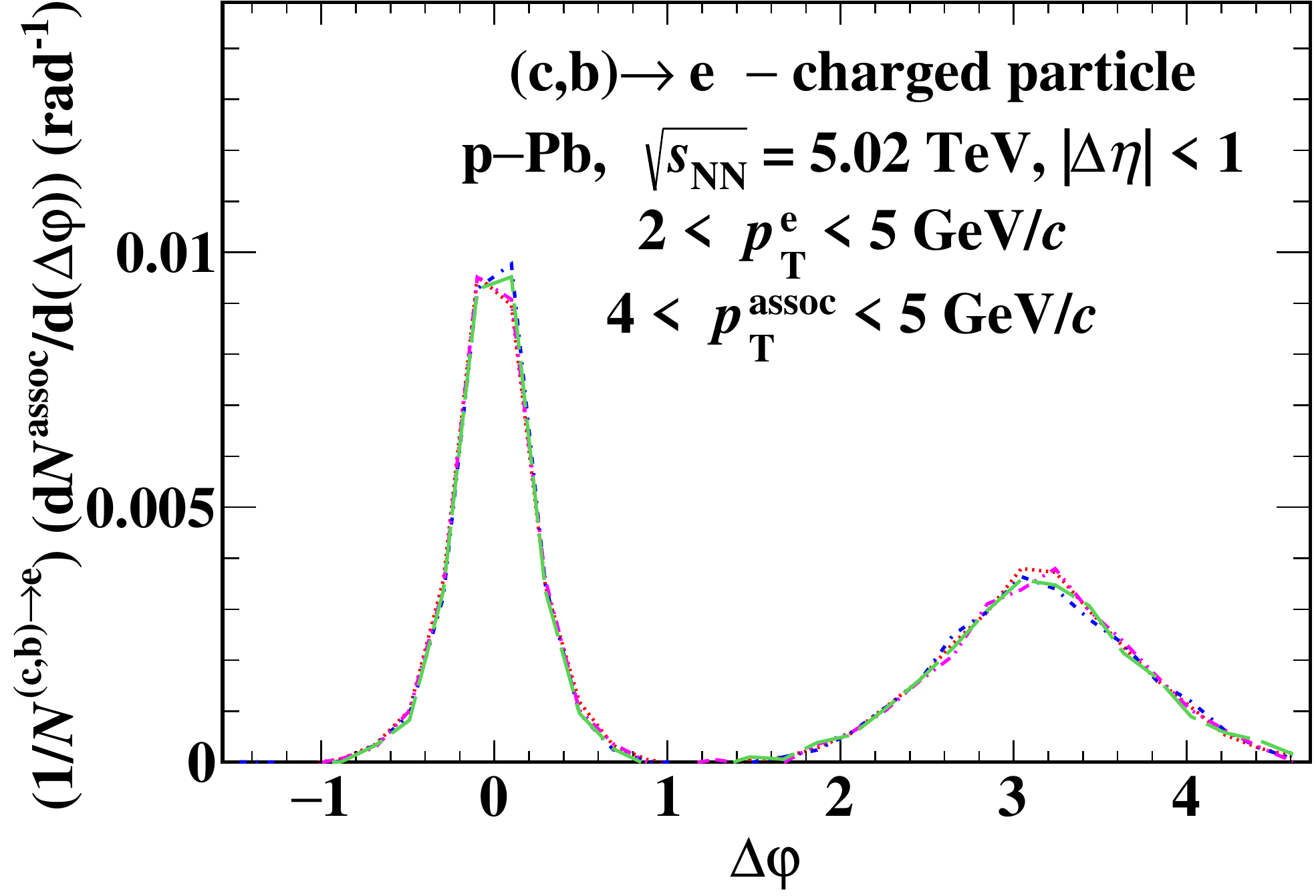}\\
\includegraphics[scale=0.40]{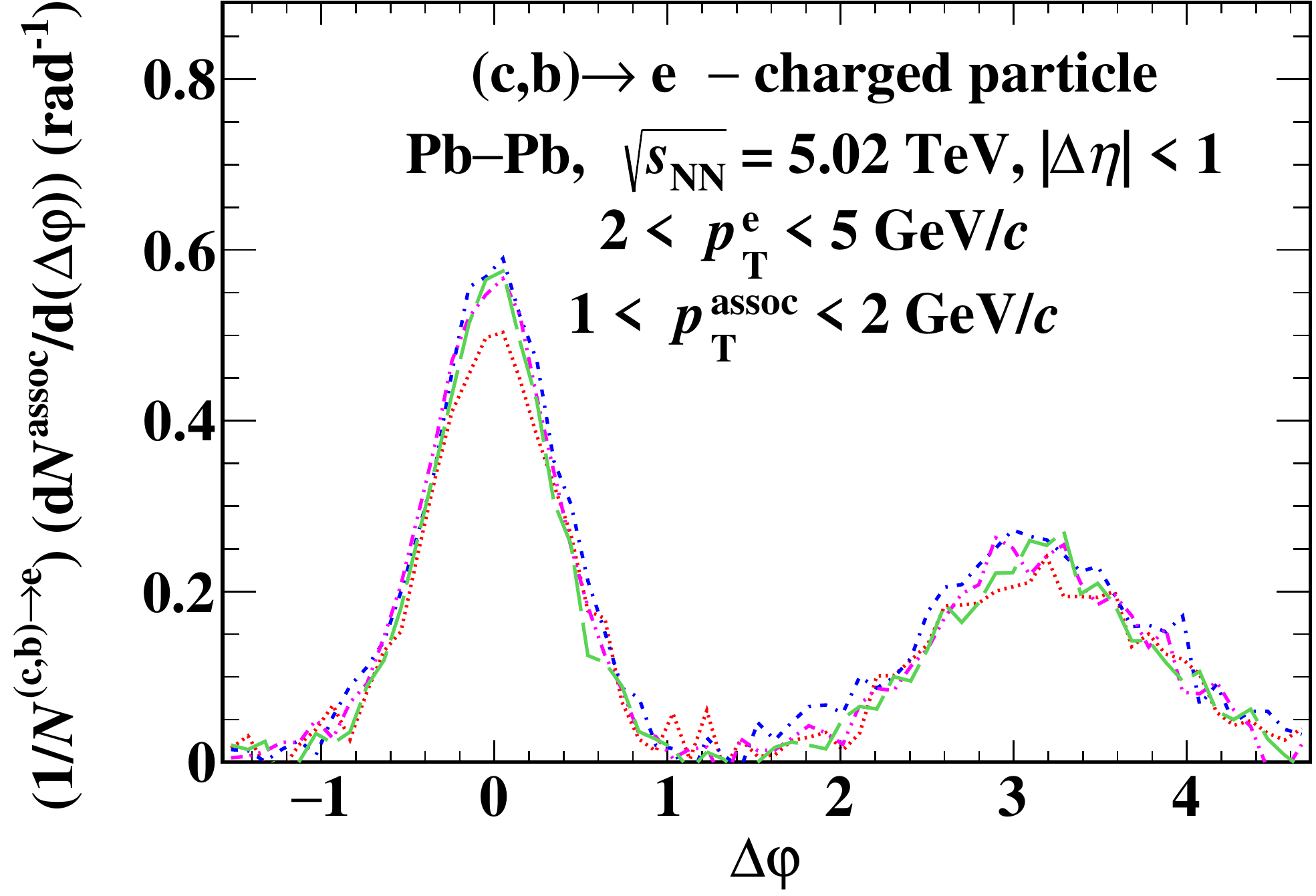}
\includegraphics[scale=0.40]{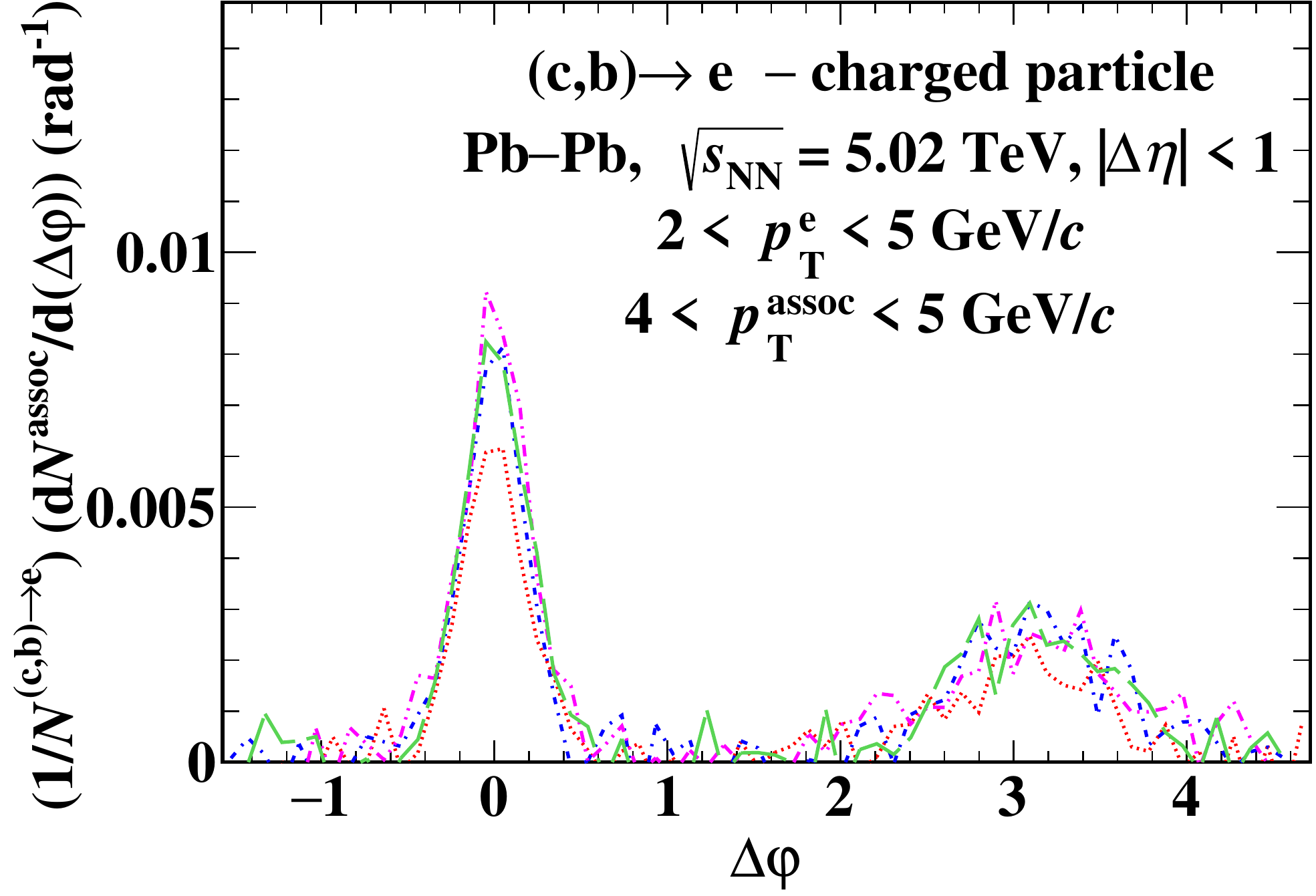}\\
\caption{(Color online) The azimuthal-correlation distribution from the PYTHIA8 for trigger \pte range $2 < p_{T}^{e} < 5$ \gev and for associated \pta ranges $1 < p_{T}^{assoc} < 2$ and $4 < p_{T}^{assoc} < 5$ GeV$/c$ in pp, p--Pb and Pb--Pb collisions at \sqsn = 5.02 TeV.}
\label{delphi2_5}
\end{figure*}

The CR mechanism of hadronization can be further explored by investigating the string topology between the partons. In the Leading Color (LC) approximation, quarks and antiquarks connected by a colored string have a unique index. This ensures a fixed number of colored strings, such that no two quarks (antiquarks) can have the same color. The same applies to gluons, represented by a pair of colored strings. This model is extended to non-LC topologies or Beyond-LC (BLC), where the colored strings can form between both LC-connected partons and non-LC-connected partons. This leads to further possibilities of a string being connected to other partons of the matching index other than the LC parton. During the reconnection process, time-dilation effects are taken into account, and three modes of Color Reconnection in the BLC approximation are taken. In this study, we use PYTHIA8/Angantyr tunes for LC (MONASH 2013) and BLC (Mode0, Mode2, Mode3) CR mechanisms and study the effect of various hadronization mechanisms~\cite{Christiansen:2015yqa}. 

The Angantyr framework within PYTHIA allows the extension to proton-nucleus and nucleus-nucleus collisions, whereas standalone PYTHIA provides only proton-proton and proton-lepton collisions. Angantyr uses a Glauber model-based eikonal approximation for the number and positions of interacting nucleons and the number of binary nucleon-nucleon collisions. It includes individual fluctuations in the nucleon substructure for both target and projectile, implementing Gribov's corrections. The collisions in Angantyr can be PYTHIA8 driven (non-diffractive), which are pp-like, which means a wounded projectile nucleon can have pp-like collisions with one or more target nucleons (secondary non-diffractive)~\cite{Bierlich:2018xfw,Bierlich:2016smv}.

The production of heavy flavors in PYTHIA is based on Leading Order (LO) perturbative processes of gluon fusion ($gg \rightarrow Q\overline{Q}$) or pair annihilation ($q\overline{q} \rightarrow Q\overline{Q}$). PYTHIA also approximates certain higher-order contributions within its LO framework via flavor excitation ($gQ \rightarrow Qg$), or gluon splittings ($g\rightarrow Q\overline{Q}$) which give rise to heavy flavor production during high \pt parton showers. Studies report that the contribution from heavy flavor electrons in PYTHIA is dominated by flavor excitation and gluon splitting~~\cite{Norrbin:2000zc}~\cite{Ilten:2017rbd}.

We used PYTHIA version 8.3 and PYTHIA8+Angantyr to generate around 50 million events for pp and p--Pb collisions at \sqsn = 5.02 TeV, respectively. For Pb--Pb, approximately 5 million events were generated using PYTHIA8+Angantyr at \sqsn = 5.02 TeV. In this paper, the results are the predictions for the ALICE experiment. Therefore, the electrons from heavy flavor hadrons ($c,b \rightarrow e$) decays are selected within $|\eta|<0.6$ as trigger particles due to the acceptance of the electromagnetic calorimeter (EMCal) detector in ALICE. The trigger particles are selected from 4 to 20 GeV$/c$. In order to increase the statistics of heavy-flavor decay electrons, the hard QCD processes are turned on to enable charm and beauty quark production with the minimum phase space cut of 9 GeV$/c$, which is a safe choice for LHC energies. The number of electrons from beauty and charm hadrons is corrected using FONLL prediction~~\cite{Cacciari:1998it}~\cite{Cacciari:2001td}~\cite{Cacciari:2012ny}, as the decay kinematics and fragmentation of charm and beauty are different. The correlation distribution of heavy-flavor decay electrons is generated by correlating each heavy flavor electron to the associated particles from 1 to 7 GeV$/c$. Here, associated particles are the physical primary particles.

To validate these settings of PYTHIA, a comparison of azimuthal correlation (\delphi distribution) of prompt D-meson and charged particles with ALICE data is shown in Fig.~\ref{fig:delphi1}. In the figure, the \delphi distribution obtained from PYTHIA8 Monash tune is compared with ALICE published data for the $\sqrt{s}$ = 7 TeV in the \pt trigger 5-8 GeV$/c$ (\pte) for associate particles \pt 0.3-1 GeV$/c$ (\pta)~~\cite{ALICE:2016clc}. Here, the range of \delphi distribution is taken from 0 to $\pi$ to match with ALICE data. The pedestal (baseline) is subtracted from the generalized Gaussian function considering the physical minima around $\pi/4$ to $\pi/2$. The result from PYTHIA shows a good agreement with ALICE data which motivates us to give a prediction on heavy-flavor electron correlation with charged particles.

\subsection{Baseline estimation and near- and away-side observable extraction}

The correlation analysis is performed by correlating each heavy-flavor decay electron with its associated charged particles. In order to measure both the near- and away-side peaks with full ranges, the \delphi distribution is obtained in the range $-\pi/2 <$\delphi$< 3\pi/2$, where the near-side peak is observed at \delphi = 0, formed by the charged particle associated with the electron of high transverse momentum (\pte) particle, whereas the away-side peak appears at \delphi $=\pi$ due to back to back di-jets produced by LO processes. A flat region also appears between the peaks formed under the signal region by the uncorrelated pairs of trigger particles and associated particles. Most of the contribution in the baseline comes from the soft processes. The baseline subtraction and measurement of near- and away-side observables are performed by fitting the raw \delphi distribution (included baseline) with the von Mises function, as shown in FIG.~\ref{fig:fitdelphi}~\cite{Hamelryck,Mardia}. The function is defined as:

\begin{align}
    f(\Delta \varphi) = b + \frac{e^{\kappa_{NS}\cos{(\Delta \varphi)}}}{2{\pi}I_{0}(\kappa_{NS})} + \frac{e^{\kappa_{AS}\cos{(\Delta \varphi - \pi)}}}{2{\pi}I_{0}(\kappa_{AS})}
\end{align}

Here, $b$ is the baseline, $\kappa$ is the reciprocal of dispersion, which means it gives a measure of the concentration, $I_{0}$ is the $0^{th}$ order modified Bessel function. The mean for near- and away-side peaks are fixed to ``0" and ``$\pi$," respectively.

\begin{figure*}
\centering
\includegraphics[clip, trim=0.0cm 4.8cm 10.5cm 0.8cm, width=0.32\textwidth]{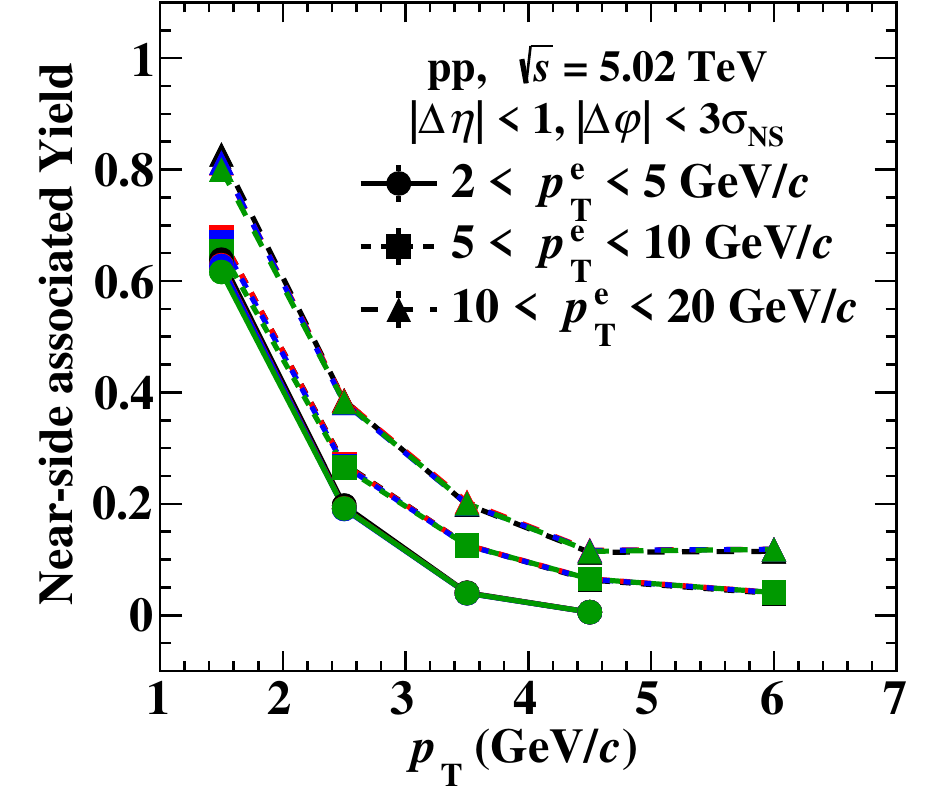}
\includegraphics[clip, trim=0.0cm 4.8cm 10.5cm 0.8cm, width=0.32\textwidth]{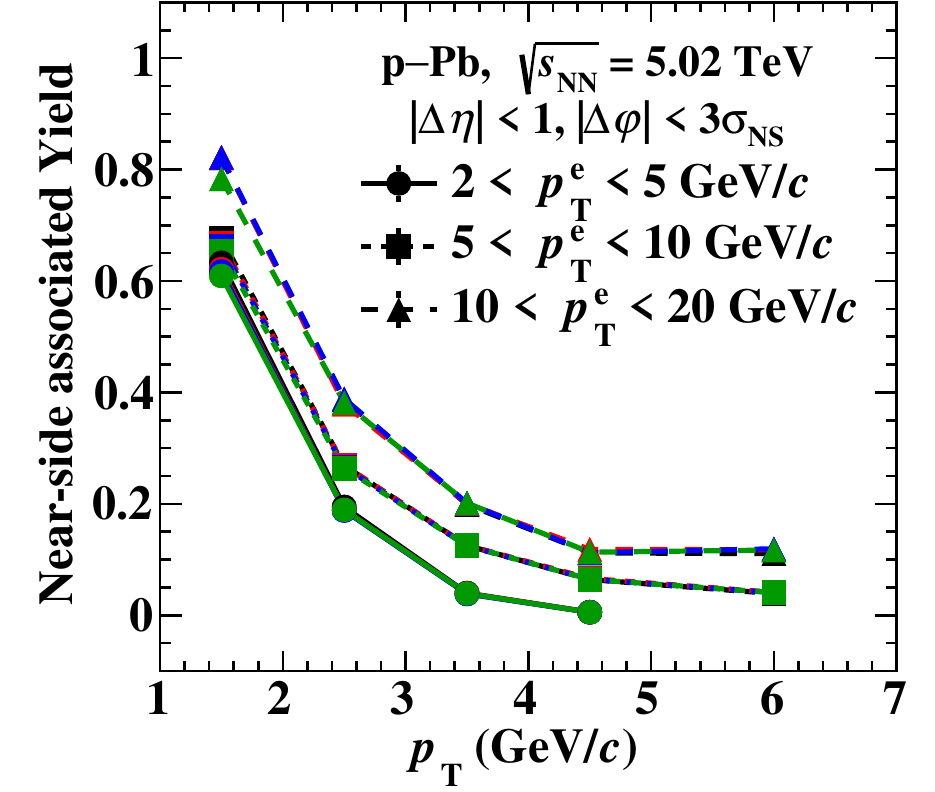}
\includegraphics[clip, trim=0.0cm 4.8cm 10.5cm 0.8cm, width=0.32\textwidth]{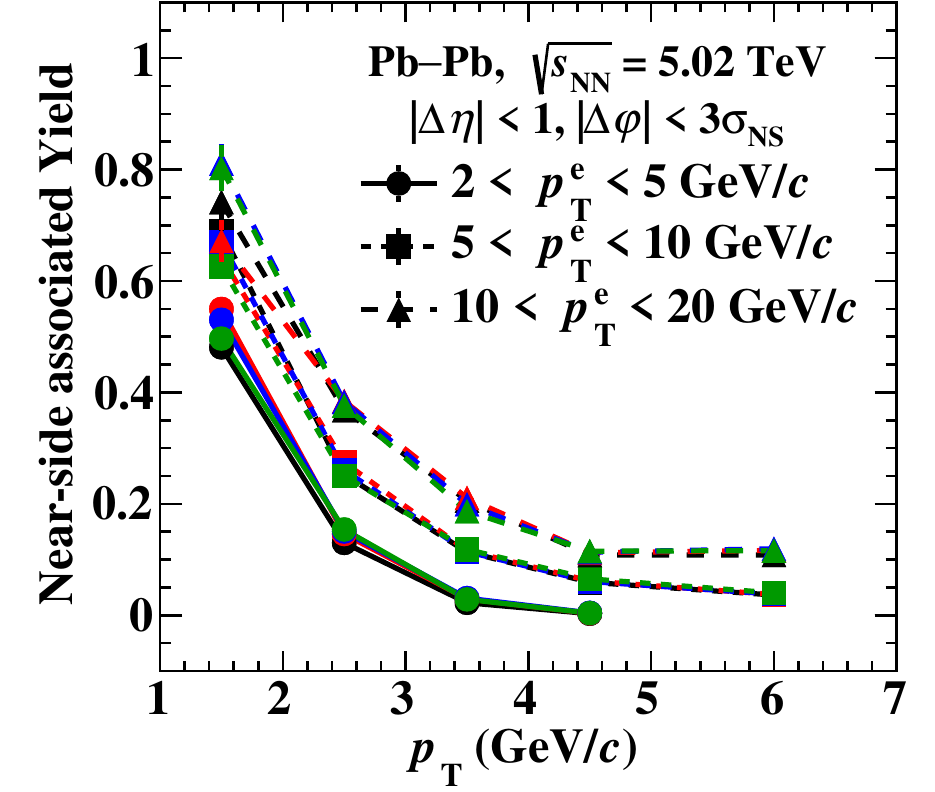}\\
 
\includegraphics[clip, trim=10.cm 4.8cm 0.5cm 0.8cm, width=0.32\textwidth]{yield_pp_trigger.pdf}
\includegraphics[clip, trim=10.cm 4.8cm 0.5cm 0.8cm, width=0.32\textwidth]{yield_ppb_trigger.pdf}
\includegraphics[clip, trim=10.cm 4.8cm 0.5cm 0.8cm, width=0.32\textwidth]{yield_pbpb_trigger.pdf}
\caption{The near- and away-side yields of correlation peaks from PYTHIA8 for different trigger \pte ranges $2 < p_{T}^{e} < 5$, $ 5 < p_{T}^{e} < 10$, and $ 10 < p_{T}^{e} < 20$ \gev for different associated \pta ranges between $1 < p_{T}^{assoc} < 7$ \gev in pp, p--Pb and Pb--Pb collisions at \sqsn = 5.02 TeV.}
\label{yields}
\end{figure*}

\begin{figure*}
\centering
\includegraphics[clip, trim=0.0cm 4.8cm 10.5cm 0.8cm, width=0.32\textwidth]{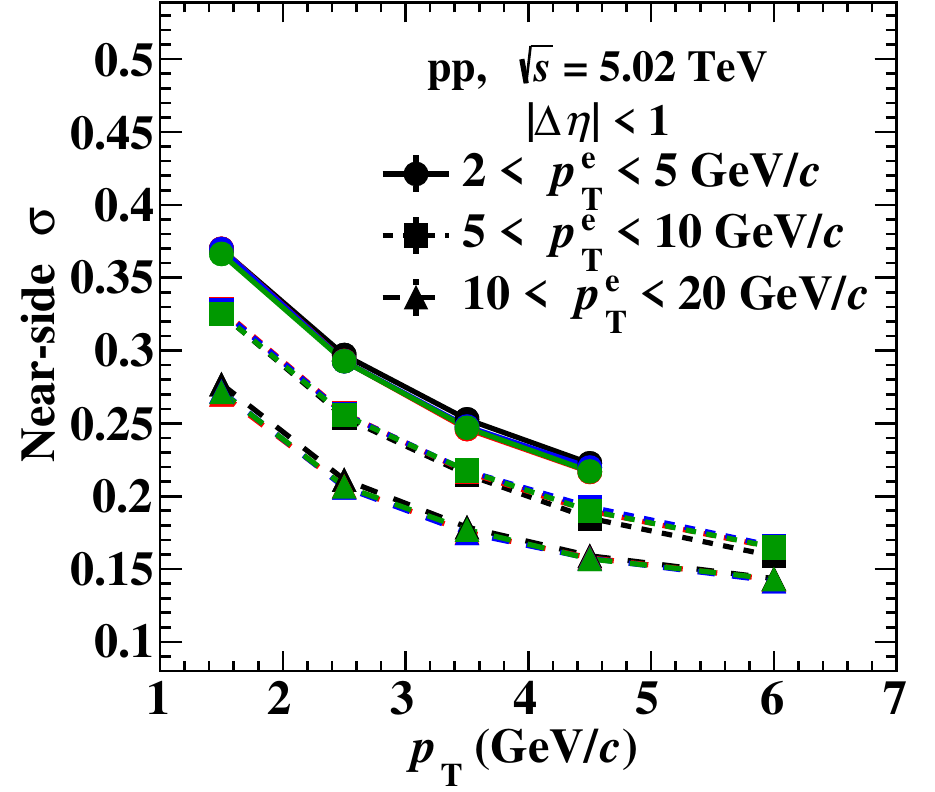}
\includegraphics[clip, trim=0.0cm 4.8cm 10.5cm 0.8cm, width=0.32\textwidth]{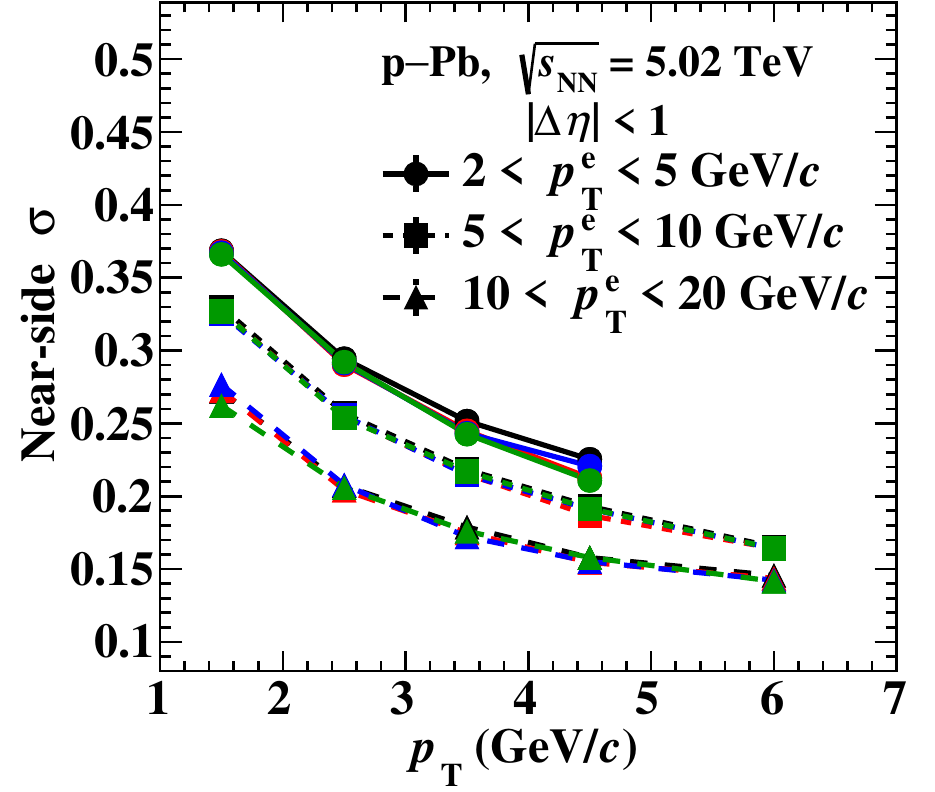}
\includegraphics[clip, trim=0.0cm 4.8cm 10.5cm 0.8cm, width=0.32\textwidth]{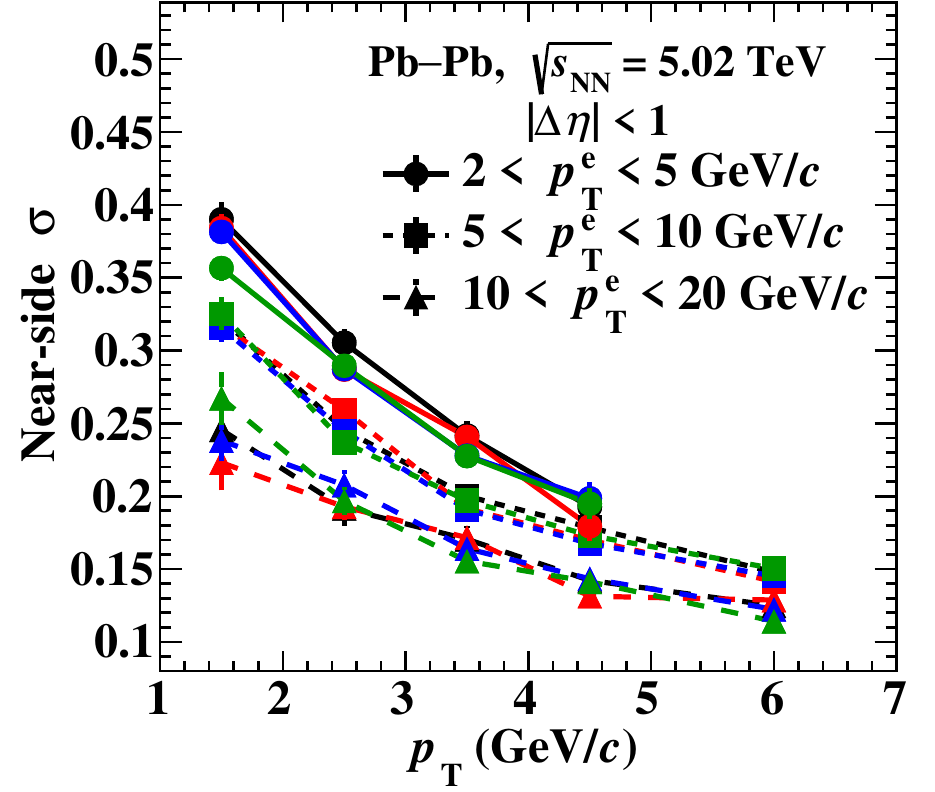}\\

\includegraphics[clip, trim=10.cm 4.8cm 0.5cm 0.8cm, width=0.32\textwidth]{sigma_pp_trigger.pdf}
\includegraphics[clip, trim=10.cm 4.8cm 0.5cm 0.8cm, width=0.32\textwidth]{sigma_ppb_trigger.pdf}
\includegraphics[clip, trim=10.cm 4.8cm 0.5cm 0.8cm, width=0.32\textwidth]{sigma_pbpb_trigger.pdf}
\caption{The near- and away-side widths ($\sigma$) of correlation peaks from PYTHIA8 for different trigger \pte ranges $2 < p_{T}^{e} < 5$, $ 5 < p_{T}^{e} < 10$, and $ 10 < p_{T}^{e} < 20$ \gev for different associated \pta ranges between $1 < p_{T}^{assoc} < 7$ \gev in pp, p--Pb and Pb--Pb collisions at \sqsn = 5.02 TeV.}
\label{sigma}
\end{figure*}

\begin{figure}
\includegraphics[scale=0.4]{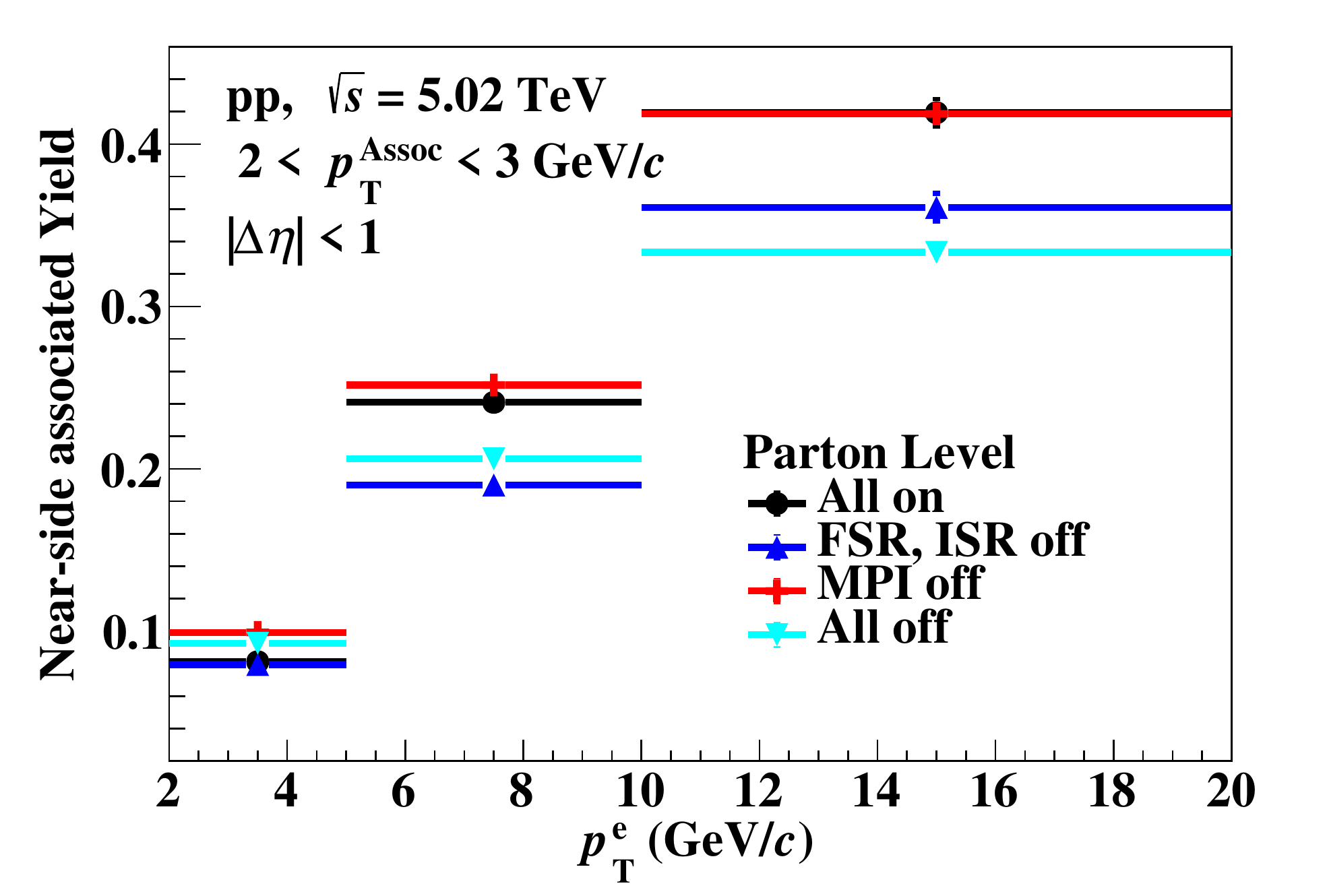}
\includegraphics[scale=0.4]{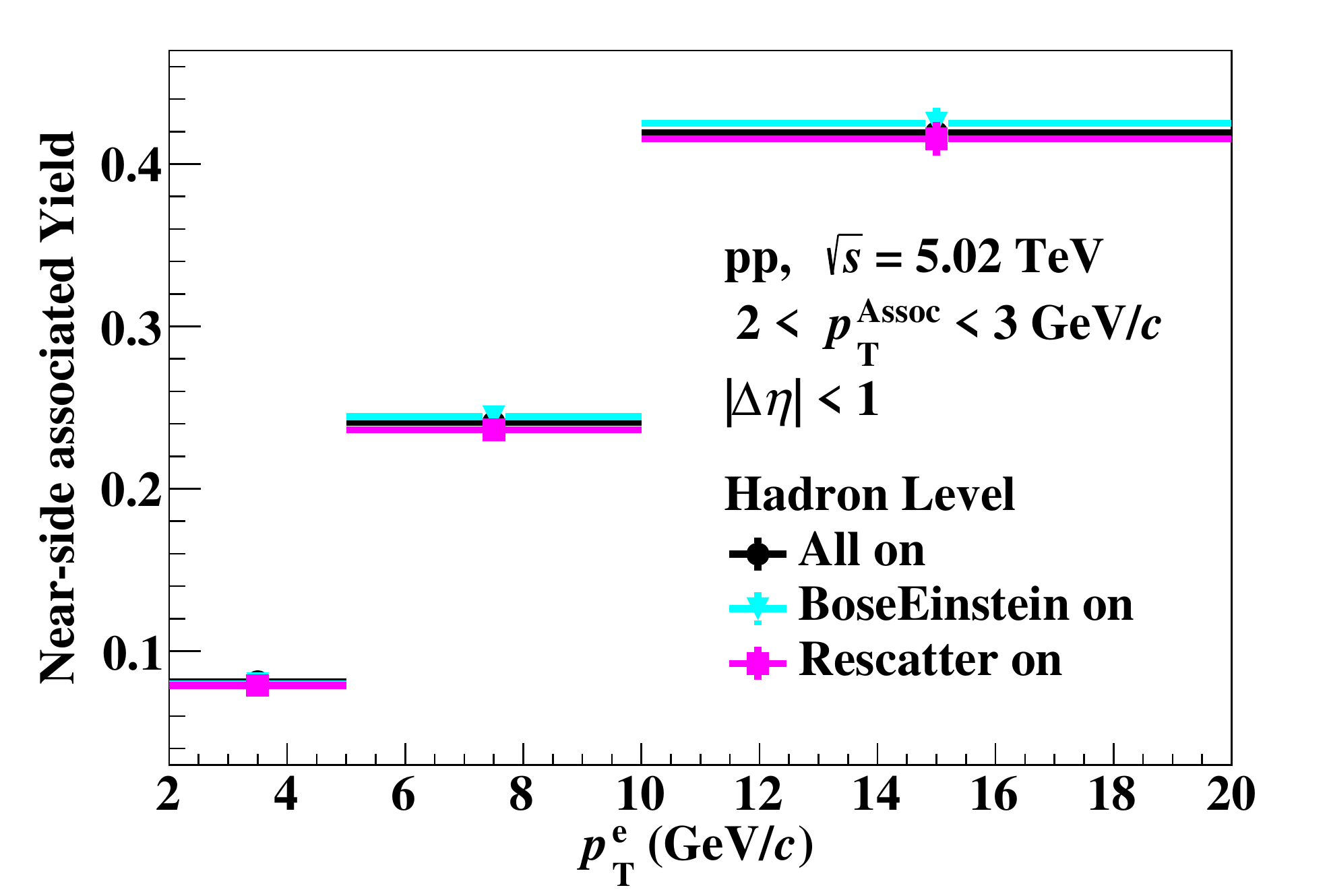}
\caption{The near-side yields of correlation peaks from PYTHIA8 for different parton level (Up) and hadron level (Down) processes for trigger \pte ranges between $2 < p_{T}^{e} < 20$ \gev and for associated \pta range $2 < p_{T}^{assoc} < 3$ \gev in pp collisions at \sqs = 5.02 TeV.}
\label{fig:Processes}
\end{figure}

Earlier, a double Gaussian, double generalized Gaussian, and generalized Gaussian + Gaussian functions, along with a constant term, were employed in these measurements to measure the near- and away-side observables as well as to estimate the baseline. But due to the triangular structure of the near-side correlation peak, the Gaussian function is not suitable as a fit function. The generalized Gaussian function is discarded as the number of free parameters is larger than that of the von Mises function, which may bias the near- and away-side observables, especially the width, as the shape parameters of the generalized Gaussian have an anti-correlation with the width of the peaks. Another advantage of the von Mises function is that it can adjust the shape according to correlation peaks, as the shape of the near-side peak is near to triangular, whereas the away-side peak is almost Gaussian. In the measurements of D meson correlation, authors used two different functions, a generalized Gaussian for near-side peak (triangular) and a Gaussian for away-side peak, where the von Mises function does not need to club with other functions.  

The near- and away-side width is estimated by measuring the sigma ($\sigma$) from the von Mises function as by the given relation:

\begin{align}
\sigma = \sqrt{-2\log\frac{I_{1}(\kappa)}{{I_{0}(\kappa)}}}
\label{eqsigma}
\end{align}

Here, $I_{0}$ and $I_{1}$ are the modified Bessel function of $0^{th}$ order and $1^{st}$ order, and $\kappa$ is measured by the von Mises function fit parameter.

The error in the width (d$\sigma$) is propagated by the relation:
\begin{align}
    \rm{d}\sigma = \frac{1}{\sigma} \times \left(\frac{I_{1}}{I_{0}} - \frac{I_{0}}{I_{1}} + \frac{1}{\kappa} \right) \rm{d}\kappa
\label{eqsigmaerr}
\end{align}

Where d$\kappa$ is the uncertainty in $\kappa$, obtained by von Mises function fitting.

In this work, we are presenting the \delphi distribution, near- and away-side yields and widths ($\sigma$) in three different \pte intervals corresponding 4-7 GeV$/c$, 7-10 GeV$/c$ and 10-20 GeV$/c$ with five \pta intervals corresponding 1-2, 2-3, 3-4, 4-5 and 5-7 GeV$/c$. The \delphi distribution obtained within $|$\deleta$|< 1$. range. A condition \pta $<$ \pte is applied while correlating the particles to avoid the double-counting of trigger electrons in correlation. These results are obtained with three different tunes of color reconnection along with the default Monash tune.

\section{Results}
\label{results}

The shape and height of the correlation peaks can be compared in pp, p--Pb, and Pb--Pb collisions to provide information about the possible system size dependence on modification of jet fragmentation. On the away-side, it reflects the survival probability of recoil partons while passing through the medium. It can be seen in Fig.~\ref{delphi2_5} that there are no significant differences among different color reconnection tunes in pp and p--Pb collisions; however, a small increment of peak height is observed in Pb--Pb collisions with BLC tunes. This might be because an additional junction was added to BLC tunes, showing the effect at high-density strings in Pb-Pb collisions. However, more study is required in this direction to make a strong claim. It is observed that the particles associated with the high \pte have higher peaks compared to low \pt trigger particles. Also, the peaks are narrower for the high \pte particle due to the initial boost. The difference between the correlation pattern can be quantified more efficiently by comparing near- and away-side yields and widths.\par
The near- and away-side width of \delphi distribution peaks are obtained for all the tunes with different triggers and associated \pt intervals, as shown in Figs.~\ref{sigma} for pp, p--Pb, and Pb--Pb collisions. By observing all the figures, it is clear that for each \pte bin, widths decrease as increasing associate particles \pt, which is reflected by the decreasing of broadness. On the other hand, peaks associated with high \pte particles have lower widths than low \pte particles due to the initial boost in the transverse direction. Different color-reconnection tunes are not showing significant changes in width, and the spreads of the widths due to the various tunes are treated as a band of systematic uncertainties.

Similarly, yields are extracted for the near- and away-side peaks. The yields are measured by the bin counting method within the three sigma ($< \rm{3}\sigma$) region from the mean value of the peaks. The $\sigma$ for the concerned peak is obtained by using eq~\ref{eqsigma} with the help of the von Mises function. The near- and away-side yields for the different \pte are shown in Figs.~\ref{yields} for pp, p-Pb, and Pb-Pb, respectively. It is observed that a high \pte particle shows a higher yield compared to a low \pte particle. This is expected as the available energy to fragment into associate particles is more prominent in high \pte particles.

Moreover, the difference in charm and beauty fragmentation could affect the yields of low and high \pte particles. The yields are decreasing towards higher \pta intervals, suggesting that fragmentation into low \pta particles is higher than high \pta particles due to the production cross-section. As the process of heavy quark fragmenting to heavy flavor hadrons is very rare, the emission of high \pt associated particles becomes limited, and most of the accompanying associated particles are softer. By comparing these yields in different systems, it is observed that the results from pp and p-Pb are consistent with each other, which is also seen in the D-meson and charged particle correlation performed by ALICE experiment~~\cite{ALICE:2016clc}~\cite{ALICE:2021kpy}. The D-h correlation measurement performed by the ALICE experiment does not show any deviation in pp and p-Pb collision results, which suggests that there is no major modification in the fragmentation due to the cold nuclear matter effect. We see the same result by using PYTHIA8 Angantyr. In contrast, yields from Pb-Pb are slightly lower, especially for low \pte particles. It must be noted that the suppression of yields (jet quenching) in Pb-Pb is due to MPI+CR and higher particle density, as a thermalized medium is not implemented in the PYTHIA8 Angantyr model. 

Further, results are obtained for different partonic and hadronization processes and compared with themselves. It provides a detailed view of the correlation function from the hard-scattering outgoing partons and their hadronization. In fig.~\ref{fig:Processes}, the near-side yields are obtained from the bin-counting method using the fit function discussed above for both parton level and hadron level processes. The top figure shows the comparison between different partonic processes, i.e., ISR, FSR, and MPI. Before hard scatterings occur, partons from the incident protons beams can radiate gluons in the initial-state radiation (ISR) process. Similarly, outgoing partons from hard-scattering processes can produce a shower of softer particles via a final-state radiation (FSR) process. Since hadrons are composite objects, more than one distinct hard-parton interaction can occur in a pp collision, and proton remnants can also scatter again on each other. Such processes are called multi-parton interactions (MPI) and are responsible for producing a large fraction of the particles. Heavy quarks in PYTHIA can occur not only from the first hard (hardest) scattering but also from hard processes in the various MPI occurring in the collisions, ordered with decreasing hardness~~\cite{ALICE:2015ikl}.

It is observed that the near side yields using all partonic processes on (default) are similar to the yields for MPI off, especially at higher $p_{\rm T}$. This is because particles produced from MPI are uncorrelated to the trigger particle; hence it contributes to the baseline. A significant decrease in yields is seen while switching off to ISR and FSR processes, as higher momentum particles contribute to more collinear particle production with these processes. This points towards a relevant role of hadronization in shaping the correlation peaks in the absence of these processes. Switching MPI off with these processes (All off) has no significant difference. The difference which we are seeing at high \pte could be due to fluctuation. In the bottom figure, different hadron level processes are shown, i.e., Bose-Einstein (BE) effect and Rescattering effect~~\cite{Fialkowski:2000iv}~\cite{Lonnblad:1997kk}~\cite{Ferreres-Sole:2018vgo}. In the phenomenological Lund Model, the BE effect is approximated by a semi-classical momentum-dependent correlation function, which effectively acts as an attractive force between two mesons. The BE class in PYTHIA performs shifts of momenta of identical particles to provide a crude estimate of BE effects. In the rescattering phenomena, it is assumed that the hadrons produced can scatter against each other on the way out before the fragmenting system has had time to expand enough that the hadrons get free. This is happening in parallel with rapid decays. It is interesting to see that no significant impact of the hadronization processes is observed in the yields. It is to be noted that in this figure, "All on" means all the default hadronic processes are on; however, BE and Rescatter are off. 
\label{result}

\section{Summary}
 \label{sum}
 In this work, we attempt to study the heavy-flavor decay electron and charged particle correlation in pp, p--Pb, and Pb--Pb collisions at $\sqrt{s_{\rm NN}}$ = 5.02 TeV using the Angantyr model incorporated in PYTHIA8. In this article, we are studying the fragmentation via heavy-flavor decay electron and the modification of fragmentation in p--Pb and Pb--Pb systems. It is to be noted that the Angantyr model is an extrapolation of pp collision into pA and AA collisions without a thermalized medium or collectivity. In this regard, the primary observations of this work are summarised below:
 
 \begin{itemize}
     \item A new fit function, named von Mises function used to fit the correlation peaks and extraction of yields and widths of near- and away-side peaks.
     
     \item No major effects are observed with different color-reconnection schemes. Color-reconnection schemes are used as systematic uncertainties.
     
     \item Effect of the initial boost on the correlation peaks is observed. Peaks are narrower for high \pte particles than lower \pte, whereas yields associated with high \pte particles are larger than lower \pte particles.
     
     \item No modification in heavy flavor fragmentation is observed in the p--Pb collision system similar to D-meson and charged hadron correlation performed on ALICE data. It suggests that the fragmentation function has no dependency on cold-nuclear matter effects.
     
     \item A small jet-quenching is seen in Pb--Pb collisions, probably due to MPI+CR and higher multiplicity compared to a small system. 
     
     \item MPI has no significant effect on fragmentation, as heavy quarks in PYTHIA can originate on different MPI stages based on the ordering of hard processes.
     
     \item Associated yields are significantly increased by ISR and FSR effects, as these radiations contribute to more collinear particle production. 
     
     \item There are no significant modifications in fragmentation due to hadron-level processes, i.e., BE effect and rescatter effect. This suggests that associated yields per trigger particle are mainly generated by parton fragmentation.  
     
 \end{itemize}
 
\section*{Acknowledgements}

R.~S. acknowledges the financial support (DST/INSPIRE Fellowship/2017/IF170675) by the DST-INSPIRE program of the Government of India. S.~K.~K. acknowledges the financial support provided by the Council of Scientific and Industrial Research (CSIR) (File No. 09/1022(0051)/2018-EMR-I), New Delhi. R.~S. is also grateful to the authors of PYTHIA8 and PYTHIA8 Angantyr.

\end{document}